\documentclass{natu}

\usepackage{subfigure}
\usepackage{wrapfig}
\usepackage{textcomp}
\usepackage{graphicx}
\usepackage{amssymb}
\usepackage{amsmath}
\usepackage{bm}
\usepackage{amsmath, amsthm, amssymb}
\usepackage{color}

\title{Signature of a topological phase transition in the Josephson supercurrent through a topological insulator}

\author{V.~Orlyanchik$^{1}$, M. P. Stehno$^{1}$, C. D. Nugroho$^{1}$, P. Ghaemi$^{1}$, M. Brahlek$^{2}$, N. Koirala$^{2}$, S. Oh$^{2}$, and D. J. Van Harlingen$^{1}$}
\begin{document}

\maketitle

\begin{affiliations}
 \item Department of Physics, University of Illinois at Urbana-Champaign, Urbana, IL 61801, USA
 \item Department of Physics and Astronomy, Rutgers, The State University of New Jersey, Piscataway, New Jersey 08854, USA
\end{affiliations}

\begin{abstract}
Topological insulators (TIs) hold great promise for realizing zero-energy Majorana states in solid-state systems. Recently, several groups reported experimental data\cite{PhysRevLett.109.056803, PhysRevLett.110.186807, Appll, sreport, NatMat.11.417,cihan} suggesting that signatures of Majorana modes in topological insulator Josephson junctions (TIJJs) have -- indeed -- been observed. To verify this claim, one needs to study the topological properties of low-energy Andreev-bound states (ABS) in TIs of which the Majorana modes are a special case. It has been shown theoretically that topologically non-trivial low-energy ABS are also present in TIJJs with doped topological insulators up to some critical level of doping at which the system undergoes a topological phase transition\cite{PhysRevLett.107.097001, PhysRevLett.109.237009, PhysRevB.87.035401}. Here, we present first experimental evidence for this topological transition in the bulk band of a doped TI. Our theoretical calculations, and numerical modeling link abrupt changes in the critical current of top-gated TIJJs to moving the chemical potential in the charge-accumulation region on the surface of the doped TI across a band-inversion point. We demonstrate that the critical-current changes originate from a shift of the spatial location of low-energy ABS from the surface to the boundary between topologically-trivial and band-inverted regions after the transition. The appearance of a decay channel for surface ABS is related to the vanishing of the band effective mass in the bulk and thus exemplifies the topological character of surface ABS as boundary modes. Importantly, the mechanism suggest a means of manipulating Majorana modes in future experiments.
\end{abstract}



In small-gap semiconductors strong spin-orbit interactions may cause an inversion between valence and conduction bands generating a new class of insulators which is called topological insulator~(TI)\cite{NatPhys.5.378, RevModPhys.82.3045, RevModPhys.83.1057}. One intriguing feature of these systems is the emergence of gapless (i.e. metallic), spin-momentum-locked states on the interface with ordinary insulators, e.g. a region with topologically-trivial bandstructure or vacuum. There has been growing interest in the properties of ordered phases of these helical states. Although, in two spatial dimensions, fluctuations prohibit spontaneous symmetry breaking\cite{PhysRevLett.17.1133}, new phases may be induced by the proximity effect when magnetic or superconducting materials are brought into contact with TIs. Coupling TIs and conventional superconductors (SC) introduces helical surface states with superconducting pair-correlations and is of particular interest in the present context. It was argued that in-gap vortex modes\cite{caroli}, and Andreev states in $\pi$-Josephson junctions of conventional superconductors which are coupled by TI surface states may carry a zero-energy Majorana mode\cite{PhysRevLett.100.096407}, a fermionic mode which is its own antiparticle\cite{NatPhys.5.614}. Early models considered an idealized scenario in which helical states existed only on a single, isolated surface of the TI, and the bulk of the material was perfectly insulating. As most of the initially discovered TIs were not true insulators in the bulk, it seemed difficult to realize the proposal in experiments. Later it was shown theoretically that topologically protected zero-energy modes exist also on the surface of doped, superconducting TIs up to a critical level of doping. At this doping level, a topological phase transition occurs in the superconducting TI\cite{PhysRevLett.107.097001,PhysRevLett.109.237009,PhysRevB.87.035401}. 
Several groups demonstrated supercurrent transport in SC-TI-SC hybrid structures\cite{NatMat.11.417, ZhangInducedSCinBiSe, MorpurgoBiSe, PhysRevLett.109.056803} as well as the ability to control the chemical potential and the magnitude of supercurrent in a TI by electrostatic gating\cite{MorpurgoBiSe, NadyaBiSe}. Furthermore, it was argued\cite{NatMat.11.417,NadyaBiSe} that the majority of the supercurrent is carried by a set of low-energy Andreev bound states located on the surface of the TI.

Recently, a new generation of high-quality, bismuth-selenide TI material became available for which it was shown that only a few quintuple layers close to the surface contribute to electrical transport\cite{ThicknessIndependentTRansport}. It was then concluded that the electric current is carried by the helical surface states and by carriers in a charge-accumulation region close to the surface of the TI where the conduction band bends down and crosses the chemical potential. As a result, a 2-dimensional electron gas (2DEG) layer is formed below the surface whereas the chemical potential in the bulk of the TI remains in the gap of the TI. We incorporate this material into a Josephson device which provides us with a unique opportunity to study the r\^ole of low-energy Andreev states and band topology in supercurrent transport as we are able to adjust the chemical potential of the 2DEG layer within the conduction band and in the gap. 

Here, we report measurements of the {D}{C} Josephson effect in SC-TI-SC devices and focus on the dependence of the critical current ($I_{C}$) on the voltage that is applied to a top-gate which is used for  electrostatic gating. Unlike the normal state resistance, which is a smooth function of gate voltage throughout, the Josephson current exhibits an abrupt drop followed by a gradual decrease as the chemical potential is lowered. We attribute the sharp change in $I_{C}$ to a topological transition in the conduction band of the charge accumulation region which shifts the spatial location of low-energy Andreev-bound states pinned on the boundary between parts of the sample with topologically-trival and band-inverted bandstructure.


The samples in this study were fabricated from high-quality Bi$_{2}$Se$_{3}$ thin-films grown by molecular-beam epitaxy on sapphire (Al$_{2}$O$_{3}$) (0001) substrates\cite{GrowBiSE1, ThicknessIndependentTRansport}. The thickness of the films varied between 6~QL and 60~QL. To pattern planar Josephson junctions, we used standard electron-beam lithography techniques. The length of the junction was set by the separation between two sputtered Nb electrodes and varied between $90$~nm and $120$~nm. The junction width was defined by dry-etching and ranged between 0.1~$\mu$m and 1~$\mu$m.  The top-gate was fabricated by evaporating 70~nm of Au on top of the 40~nm-thick dielectric layer of ALD-grown alumina (Al$_{2}$O$_{3}$) or hafnia (HfO$_{2}$). Most of the measurements were carried out in a dilution refrigerator equipped with a 10 Tesla superconducting magnet and a base temperature of 8 mK.  The data, we present here, were obtained on a junction of 0.1~$\mu$m length, 0.4~$\mu$m width, and a thickness of 15~QL and are representative for all 15 measured samples.

\begin{figure}[h!]
    {\includegraphics[width=7.5cm]{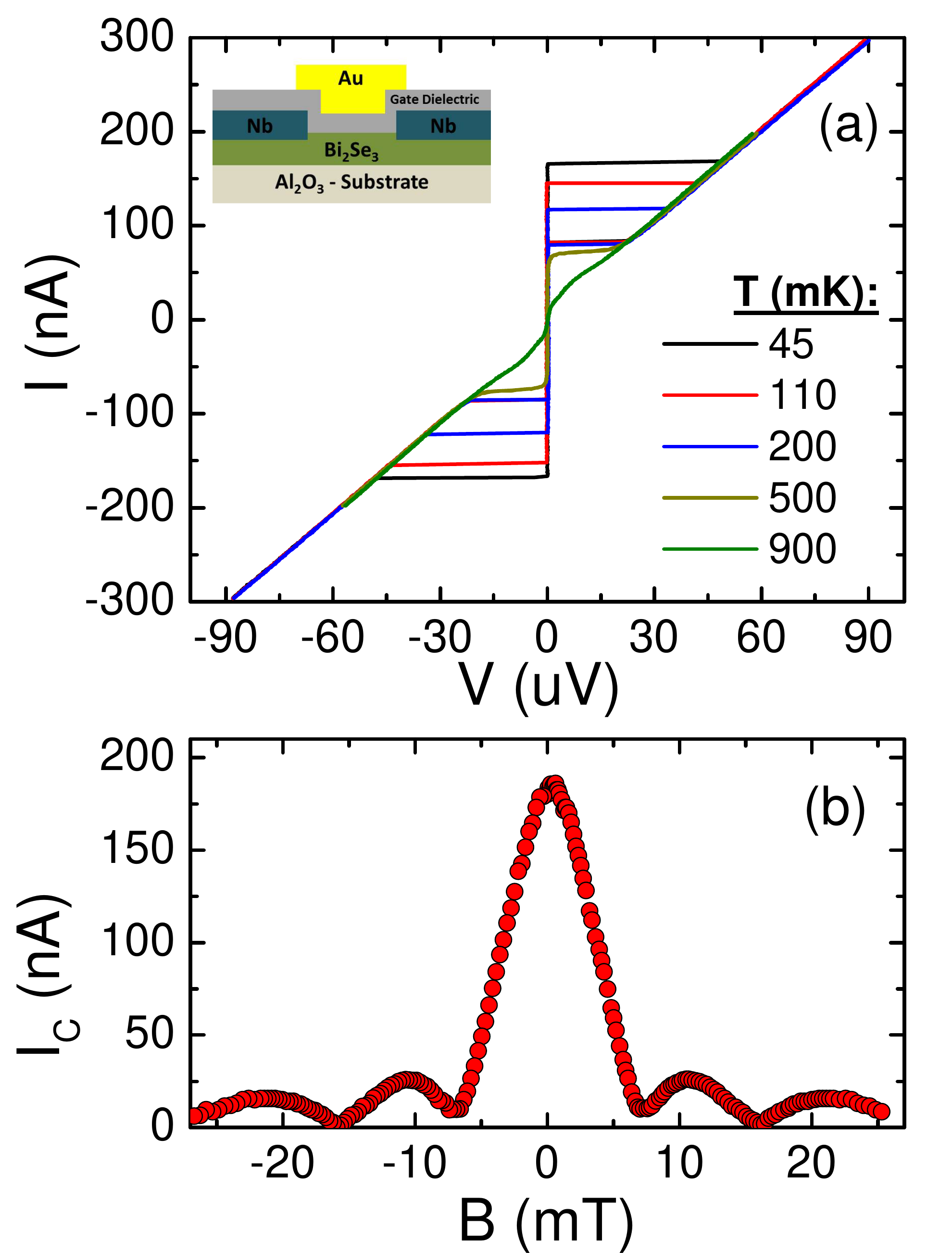}}
    \caption{(a)~The current-voltage characteristics of the Josephson junction at different temperatures. Inset:~A schematic representation of the junction cross-section. (b)~The dependence of the critical current on perpendicular magnetic field exhibits a Fraunhofer-like pattern.}
     \label{fig:I-V}
\end{figure}

The current-voltage (I-V) characteristics of a Nb-Bi$_{2}$Se$_{3}$-Nb junction is shown in Fig.~\ref{fig:I-V}(a) for different sample temperatures. Here, we define the critical current ($I_{C}$) of the device as the current at which a finite voltage drop develops between the two superconducting electrodes. A maximum Josephson supercurrent of $I_{C}^{max}=190$~nA  was measured at temperature $T=45$~mK. Other devices showed values of $I_{C}^{max}$ between 10~nA and 300~nA scaling with the width of junction. The hysteresis in the low-temperature I-V characteristics is consistent with electron heating which develops after switching the junction into the resistive state\cite{HysteresisSNS}. As shown in Fig.~\ref{fig:I-V}(b), a Fraunhofer-like pattern is generated by applying a perpendicular magnetic field which modulates the critical current in the device. While the general shape of the curve is typical for Josephson devices, the details specific to SC-TI-SC junctions have been discussed elsewhere\cite{cihan}.

Having verified the presence of Josephson coupling in our devices, we turn to the doping dependence of the Josephson current which was studied by electrostatic depletion of charge carriers in the junction. The intrinsic n-doping that we observed in our devices is commonly attributed to selenium vacancies forming in MBE-grown Bi$_{2}$Se$_{3}$ films shortly after the growth is completed\cite{EviromentalDisorder1,EviromentalDisorder2}. The excess electronic density raises the Fermi energy in the device and can be removed effectively by applying a negative voltage to the top-gate of the junction (cp. device schematics in the inset of Fig.\ref{fig:I-V}(a)).

\begin{figure}[h!]
    {\includegraphics[width=9.5cm]{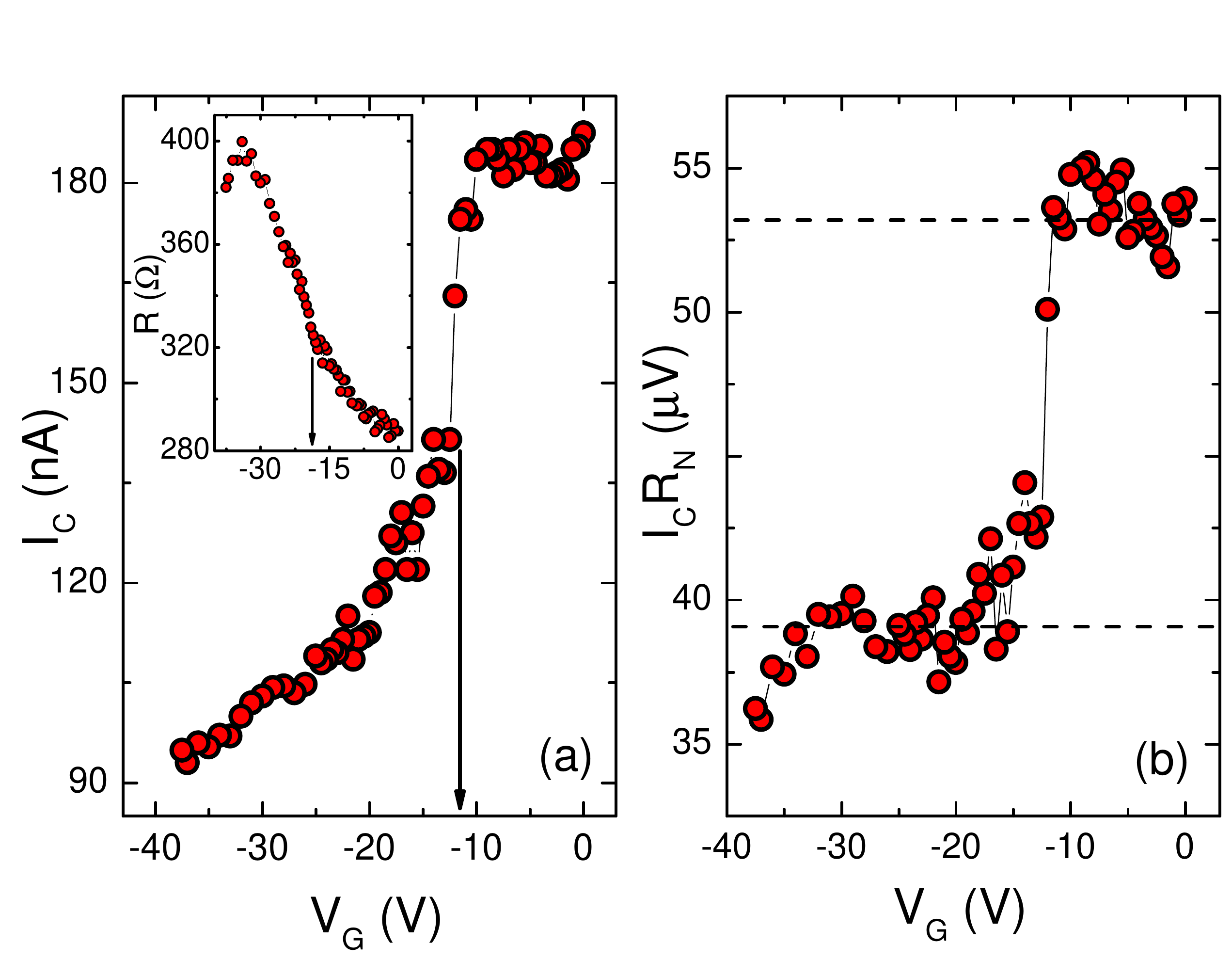}}
    \caption{(a)~The critical current as a function of the gate voltage measured at $T=50$~mK. Inset:~Dependence of the normal state resistance of the junction on gate voltage. (b)~The variation of the product of the critical current and normal state resistance ($I_{C}R_{N}$) measured as a function of gate voltage. The dashed lines emphasize two values of $I_{C}R_{N}$ above and below the transition region.}
     \label{fig:Ic-Vg}
\end{figure}

A change in charge-carrier density affects resistance and critical current of the junction in different ways. As shown in Fig.~\ref{fig:Ic-Vg}(a), applying a negative gate voltage reduces the critical current in a non-monotonic manner. The initial, gradual reduction of $I_C$ is followed by a rather abrupt drop. This rapid change in critical current takes place in a narrow region of gate voltages, $\Delta V_{G}<1~$V, which we call transition region and mark it by its center value, the critical gate voltage $V_{G}^{C}\approx -12$~V. The transition region is trailed by another gradual and featureless decrease of $I_{C}$. This highly non-monotonic behavior of $I_{C}(V_{G})$ is reproducible. It was observed in multiple, consecutive gate voltage scans for each individual Josephson junction, and it was present in all measured devices (15 Josephson junctions). The value of $V_{G}^{C}$ varies accordingly with thickness and dielectric constant of the gate dielectric (Al$_{2}$O$_{3}$ or HfO$_{2}$). In contrast, the abrupt change in $I_{C}$ was \textit{not} accompanied by a fast variation in the normal state resistance of the junction, which is a smooth function of $V_G$ (see inset in Fig. \ref{fig:Ic-Vg}(a)). Moreover, as shown in Fig. \ref{fig:Ic-Vg}(b), the product of critical current and normal state resistance, $I_{C}R_{N}$, which reflects the characteristic energy of the Josephson coupling, shows a distinct step at $V_{G} \approx V_{G}^{C}$.

We propose that the abrupt change in the critical current is a manifestation of a topological phase transition in the charge-accumulation region of the sample. Below, we argue that a considerable fraction of the supercurrent is carried by Andreev-bound states (ABS) which are pinned on the interface ($\it{boundary}$ ABS) between the topologically-trivial region and the band-inverted insulating bulk of the TI. Electrostatic gating changes the bending of the conduction band at the surface and, thus, moves the location where the chemical potential intersects with respect to the bottom of the band. This allows the 2DEG layer to transition between a topologically-trivial and a band-inverted phase.

\begin{figure}[h!]
    {\includegraphics[width=13.5cm]{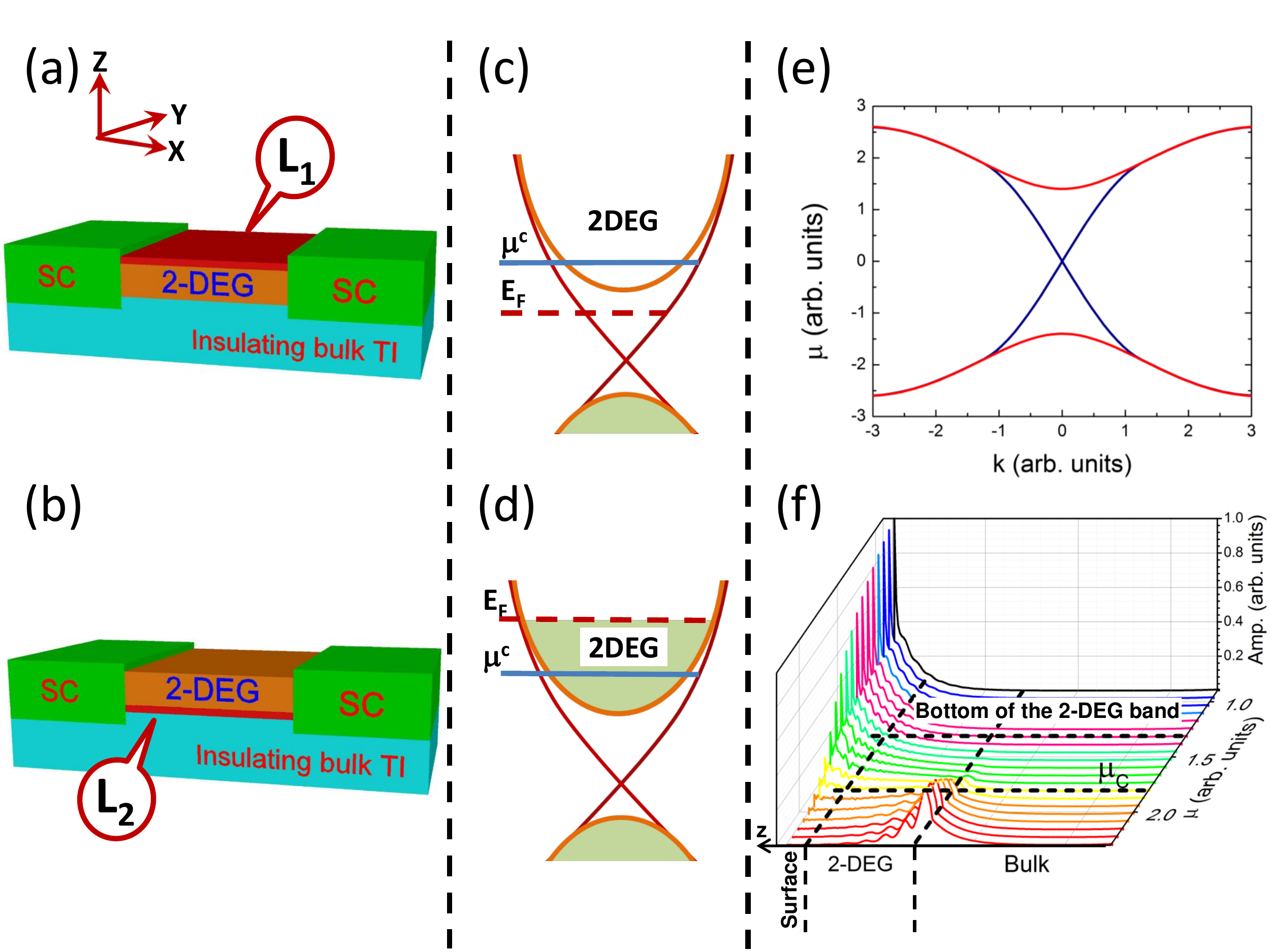}}
    \caption{(a) and (b) show a schematic representation of the location of topological Andreev-bound states corresponding the positions of the chemical potential in~(c) and~(d). (e)~The band structure used in numerical calculations. (f)~The evolution of the spatial location of low-energy Andreev-bound states as a function of the chemical potential assuming the band structure in~(e).}
     \label{fig:Fig3}
\end{figure}

To demonstrate the mechanism behind the topological transition, we developed a model based on the concepts laid out in Refs.\cite{PhysRevLett.107.097001, PhysRevLett.109.237009, PhysRevB.87.035401}. It was shown that low-temperature transport in MBE-grown Bi$_{2}$Se$_{3}$ films is described fully by two conductance channels, the topological surface and a quantum-confined electron gas (2DEG) which is a few quintuple layers thick. The latter results from a downward bending of the conduction band due to charge-accumulation in a few nm-wide zone below the film surface\cite{ThicknessIndependentTRansport, TranspotinBiSE1}. Schematically, we can picture the SC-TI-SC Josephson junction as a two-layered structure of doped TI material comprised of an insulating bulk (i.e. the chemical potential is in the bulk gap) and the charge-accumulation region. Due to the short charge-screening length in the TI, we assume that electrostatic gating acts only on the 2DEG layer. This allows us to adjust its chemical potential, which -- for simplicity -- is taken to be constant throughout the layer. Looking at the cross-section of the TIJJ, we identify two important boundaries: $L_{1}$ is the physical boundary of the TIJJ between the 2DEG and vacuum (or the gate dielectric in an actual device, see Fig.\ref{fig:Fig3}(a)), and $L_{2}$ is the boundary between the insulating bulk of the TI and the 2DEG (see Fig. \ref{fig:Fig3}(b)). 

Let us first assume that the chemical potential in the 2DEG layer is well within the insulating gap (i.e. a large negative voltage is applied to the top-gate), such that all 2DEG states are depopulated (see Fig.\ref{fig:Fig3}(c)). In this case, the 2DEG region is part of the insulating bulk. The proximity effect induces pair-correlations in the topologically-protected surface layer, $L_{1}$, and the supercurrent is carried exclusively by ABS close to the physical surface of the TIJJ. Since the occupation probability of ABS follows a thermal distribution, i.e. $I_C \propto - \sum_{E_n \geq 0} \frac{\partial E_n}{\partial \phi} \tanh\left({\frac{E_n}{ 2k_B T}}\right)$, at low temperatures a large fraction of the supercurrent is carried by the ABS which are lowest in energy ($E_n$ is the energy of the $n$-th ABS for a phase difference of $\phi$ across the junction, $k_B$ is the Boltzmann constant, and $T$ is the temperature). Typically, close to $\phi = \pi$, the slope $\frac{\partial E_n}{\partial \phi}$ is largest for the lowest ABS band in a TIJJ\cite{PhysRevB.86.214515}, and under appropriate conditions this band may host a zero-energy Majorana state.

By contrast, when the chemical potential enters the 2DEG conduction band, superconductivity is induced in this region as well, and a new set of ABS emerges, which extends throughout the 2DEG layer.  
The energy of 2DEG ABS exceeds that of the boundary ABS\cite{PhysRevLett.107.097001} located at $L_{1}$ (see supplementary materials) and, thus, they are decoupled from ABS on the physical surface. Notice that they do not provide a decay -- or delocalization -- channel for Majorana modes indicating that the band topology has not changed. Indeed, previous theoretical work\cite{PhysRevLett.107.097001, PhysRevB.87.035401, PhysRevLett.109.237009} suggests that the energy of ABS in a doped TI depends on the chemical potential in a nontrivial way relating to the topology of the band. In particular, the energy of the lowest ABS  is minimal and equal to the energy of the boundary ABS when the chemical potential reaches the point of a topological transition in the conduction band (i.e. the band effective mass is zero). Below, we outline the energetics of ABS close to a topological phase transition and argue that they drive the step-like change in the critical current of our TIJJs. For a detailed theoretical discussion, we refer the reader to the supplementary material with this Letter.

To study the low-energy ABS in the 2DEG, we use the low-energy effective Hamiltonian of a doped TI with $s$-wave superconducting pair-correlations:

\begin{equation}
H_\mathrm{eff}=\left[v_F\left(k_x \sigma_x+k_y \sigma_y+k_z \sigma_z  \right) \tau_x+m \left(|\textbf{k}| \right) \tau_z - \mu \right] \eta_z + \Delta\left(\cos(x) \eta_x+\sin(x) \eta_y \right)\ .
\end{equation}\label{eq.Hamiltonian}

Here, $\eta_i$ are Pauli matrices that act on superconducting particle- and hole-states, and the representations $\tau_i$ and $\sigma_i$ are chosen for orbital- and (physical) spin-degrees of freedom\cite{PhysRevLett.107.097001}. The Hamiltonian is  translational invariant along the $\hat{y}$-direction, which is in-plane and perpendicular to the current flow in the junction (see Fig.\ref{fig:Fig3}(a)), and $k_{y}$ denotes the wave vector along $\hat{y}$. The charge-carrier concentration is determined by the chemical potential in the 2DEG region which controls the accessible  Fermi momenta $\textbf{k}$ in the band structure. These enter the equations in form of the momentum-dependent effective mass term for the conduction band, $m(|\textbf{k}|)$. A simplified form of the effective mass term, which captures the essential features of the bulk band structure of TIs, is $m \left(|\textbf{k}| \right)=M-\epsilon |\textbf{k}|^2 $ with $M\cdot \epsilon>0$. As mentioned above, at lower doping levels (i.e. closer to the Dirac point where $k = |\textbf{k}|<M/\epsilon$), the effective mass has the same sign as the mass of the bulk TI, which is given by $m \left(0\right)=M$. The low-energy ABS are localized at $L_1$, the physical surface of the TIJJ. Further increasing the chemical potential decreases the magnitude of the effective mass $|m(k)|$, which vanishes when $k=\sqrt{M/\epsilon}$. Beyond this point, the mass term reverses its sign with respect to the bulk mass $M$. As derived in the supplementary material section, the energy difference between 2DEG ABS and boundary ABS at $L_1$ is given approximately by:

\begin{equation}
|\Delta E| \propto \alpha \frac{m\left(|\textbf{k}_f|\right)^2}{m\left(|\textbf{k}_f|\right)^2+v_F^2 |\textbf{k}_f|^2}\ .
\end{equation}\label{eq.deltaE}

where $\alpha >0$ is of order of $|\Delta|^2/\mu$. Hence, for $m\left(|\textbf{k}_f|\right) = 0$, the energy of 2DEG ABS is equal to the surface bound-state energy, which opens a decay channel for the ABS at $L_1$. At the same time, the 2DEG layer becomes a region with topologically-trivial bandstructure similar to an ordinary superconducting metal and, thus, provides a de-localization path for surface ABS. As we increase the chemical potential further, we move the topological boundary to $L_2$, the interface between the insulating TI and the 2DEG region~(see Fig.\ref{fig:Fig3}(b)). Here, a new set of low-energy boundary ABS, which carries a significant portion of the supercurrent, appears after the topological transition of the 2DEG layer.    

To confirm the conjecture regarding the distribution of the supercurrent within the sample, we performed a series of numerical calculations that map out the evolution of the spatial location of low-energy ABS as a function of chemical potential. We used a simple four-band model for strong topological insulators\cite{PhysRevLett.107.097001, PhysRevB.87.035401} and considered $s$-wave superconducting pairing at the mean-field level. The energy of ABS is derived using exact diagonalization (cp. supplementary material section). Figures~\ref{fig:Fig3}(f) and~(e) show -- respectively -- the amplitudes of the low-energy ABS and the band structure used in the calculations. For the chemical potential situated within the Dirac cone ($\mu < 1.4$), supercurrent transport is localized close to the surface where the amplitude of the ABS wavefunction is large in Fig.~\ref{fig:Fig3}(f). When the critical chemical potential ($\mu_{C}$) associated with the topological transition is reached, the low-energy ABS migrate abruptly to the interface between the 2DEG layer and the insulating bulk TI. Importantly, the value of $\mu_{C}=1.8$ is significantly above the bottom of the 2DEG conduction band. Increasing the chemical potential beyond $\mu_{C}=1.8$ does not move the position of low-energy ABS.

Indeed, close inspection of the $R(V_{G})$-dependence in Fig.\ref{fig:Ic-Vg}(a) reveals that the inflection point, which is associated with the depletion of the 2DEG band, occurs at $V_{G}^\mathrm{2DEG}\sim -20$~V  (marked by an arrow in the inset). This value is distinctly lower than the critical gate voltage, $V_{G}^{C}=-12$~V. Although the position of the chemical potential at which the transition takes place varies slightly between samples (likely reflecting different levels of disorder in the devices), the width of the transition is always much smaller than the distance to the band edge, i.e. $\Delta V_{G} \ll |V_{G}^\mathrm{2DEG}-V_{G}^{C}|$, which is consistent with our assumption that the transition is controlled by properties of the bulk band structure. 

The pronounced changes in the critical current across the topological transition reflect a difference in the effective transmission of individual ABS. Whereas we may assume that the SC-TI interface barrier is lower and structural disorder is less for ABS at $L_{2}$, the effective thickness of the boundary-zone is larger on this interface. All three aspects enhance supercurrent transport by ABS at $L_{2}$. This explains the observed increase in critical current as the chemical potential is shifted up in energy across the transition and exemplifies a direct consequence of band structure topology on Andreev transport. In turn, this effect suggests a control mechanism for the spatial location of low-energy ABS and, perhaps, Majorana fermions.
 
When the junction is in the dissipative state, the current is carried by ordinary quasi-particle excitations. All conduction channels are expected to contribute, and a strong dependence on the position of the topological boundary is absent. Indeed, the normal-state resistance varies smoothly with gate voltage (i.e. charge-carrier density), cp. inset in Fig.\ref{fig:Ic-Vg}(a). As a result, a step in the $I_{C}R_{N}$-product marks the crossover between the two topologically-distinct bandstructure configurations of the 2DEG region (see Fig. \ref{fig:Ic-Vg}(b)) and illustrates the difference in the effective Josephson coupling for supercurrent transport in the two respective locations of the topological boundary.


In conclusion, we presented the experimental observation of a topological phase transition in the bulk band of a doped, proximity-coupled 3-dimensional topological insulator. Hallmark of the transition is a shift in the spatial location of low-energy Andreev-bound states which follow the position of the topological boundary. In particular, we demonstrated that the charge-accumulation zone (2DEG channel) close to the surface of a doped 3-d TI can be driven through the transition by shifting the chemical potential electrostatically with a top-gate. This was registered as a jump in the magnitude of the critical current of the Josephson junction. The abrupt change occurred within the bulk band of the 2-DEG region and correlated with a sign-change of the effective mass in the TI band structure model. The transition in the bandstructure topology of the gapped, 2DEG proximity region resulted in an altered current-flow pattern due to a displacement of low-energy ABS. Consequently, the change in the effective Josephson coupling led to a sudden and unusual drop in the $I_CR_N$-product well above the band edge. Notice that in our model for the sudden drop in the critical current, it is not necessary for the bulk of the junction to be insulating. Our model is valid even when the bulk has the fermi level in the conduction band, as long as it is not above the critical chemical potential where the topological phase transition happens.
Exploration of Majorana physics in 3-dimensional topological insulators requires control over the spatial location of zero-energy Andreev states. Our result directs toward an efficient way of manipulating low-energy Andreev-bound states in 3-d TI Josephson devices.

\section*{acknowledgments}

We would like to thank Cihan Kurter, Aaron Finck, Ashvin Vishwanath, Taylor Hughes, Jeffrey Teo and Eduardo Fradkin for useful discussions. VO, MPS, CDN, and DJVH acknowledge funding by Microsoft Station-Q. For the device fabrication, we acknowledge use of the facilities of the Frederick Seitz Materials Research Laboratory at the University of Illinois at Urbana-Champaign. PG acknowledges support from NSF DMR-1064319. The work at Rutgers was supported by IAMDN of Rutgers University, NSF DMR-0845464 and ONR N000141210456.

\bibliography{gatedTI}

\eject

\begin{addendum}
\item[\large{Supplementary Materials}]

\

\textbf{Low energy Andreev states in the bulk and on the surface of doped topological insulators} 

Gapless helical surface states are considered the characteristic feature of topological insulators. Topological properties of bulk bands are difficult to observe. However, close to a transition, the Andreev-bound state (ABS) spectrum in Josephson junctions depends sensitively on the bulk-band topology which allows us to detect a clear signature of a topological transition in the band.

At wave vectors close to different time-reversal symmetric points (TRs), the Hamiltonian of bulk TI bands resembles the three-dimensional massive Dirac Hamiltonian. At the TRs, a relative change in the sign of the mass term in the Dirac Hamiltonian occurs\cite{PhysRevB.76.045302}. As a result, the minimal effective Hamiltonian for the wave vectors near the bottom of the conduction band of 3-D TIs is given by:
\begin{equation}\label{bh}
\mathcal{H}=v_F \tau_x \ {\boldsymbol  \sigma} \cdot \textbf{k} + \tau_z \ m(\textbf{k})
\end{equation}
where ${\boldsymbol  \sigma}=(\sigma_x,\sigma_y,\sigma_z)$ are the Pauli matrices acting on spin space, $\tau_x$, $\tau_y$ are the Pauli matrices acting on orbital space, and $\textbf{k}$ is the wave vector relative to the TR at the bottom of the conduction band. The momentum-dependent mass term $m(\textbf{k})=M-\epsilon |\textbf{k}|^2$, with $M\cdot \epsilon >0$, changes sign at $|\textbf{k}| =\sqrt{M/\epsilon}$. It was shown before that the topological properties of the superconducting phase of a doped topological insulator change when the chemical potential corresponds to the wave vectors at which the effective mass $m(\textbf{k})$ vanishes\cite{PhysRevLett.107.097001,PhysRevLett.109.237009,PhysRevB.87.035401}. One manifestation of this topological transition is the appearance of edge modes, e.g. the zero-energy Majorana states at the ends of a vortex passing through a doped topological insulator\cite{PhysRevLett.107.097001, PhysRevLett.109.237009, PhysRevB.87.035401, PhysRevLett.100.096407}. In this case, the bound states that extend along the vortex\cite{caroli, PhysRevLett.107.097001, PhysRevLett.109.237009, PhysRevB.87.035401} become gapless at the transition and provide a channel for coupling Majorana modes at both ends of the vortex thus allow them to annihilate. Indeed, zero-modes in the energy spectrum of the superconducting bulk were the first theoretical evidence for a topological phase transition as function of the doping level in doped TIs\cite{PhysRevLett.107.097001}. Later the result was confirmed by looking directly at the evolution of Majorana states at the end of the vortex as a function of the chemical potential\cite{PhysRevB.84.144507,PhysRevB.87.035401}. 

An analogous scenario is realized in a topological insulator Josephson junction (TIJJ). When the chemical potential is in the bulk gap of the TI, ABS are found only at the TI surface. At a phase difference of $\pi$, two of the localized Andreev states in a TIJJ are zero-energy Majorana modes\cite{PhysRevLett.100.096407}. On the other hand, when the TI is doped and the chemical potential enters the conduction band, superconductivity is induced in the bulk, and additional ABS are formed throughout the TIJJ. Similar to the topological transition in the vortex, the energy of ABS evolves with the chemical potential in the TI. 
As we move it through the bulk band of the TI, the shape of the Fermi surface changes, i.e. the Fermi surfaces is composed of different sets of \textbf{k}-states.  For particular \textbf{k}-vectors, the mass term $m(\textbf{k})$ vanishes, and the bandstructure undergoes a topological phase transition. An important signature is that the ABS spectrum becomes gapless thus decay channels for Majorana modes open, which -- at this point -- can no longer exist on the surface. 

The presence of Majorana modes on the surface and gapless modes in the bulk of TIJJs corresponds to a phase difference of $\pi$, precisely. Under simplifying assumptions, it can be shown that this condition is fulfilled in TIJJs at maximum critical current\cite{PhysRevB.86.214515}. We, however, are interested in the full evolution of low-energy Andreev states. Thus, we study the general case of TIJJs with arbitrary phase difference. In what follows, we first derive the energy of ABS on the surface of a narrow junction. Next, we find the energy of bulk ABS and show that it decreases as the magnitude of the effective mass, $|m(\textbf{k})|$, decreases. In particular, we see that -- for vanishing effective mass~$m(\textbf{k})$ -- the (finite) energy of the lowest-lying ABS on the surface will become equal to the energy of the bulk Andreev states even if the phase of the junction is not $\pi$.

When the chemical potential is in the bulk band-gap of the TI, the only gapless states are helical states localized at the boundary of the TI sample. They realize the gapless, 2-dimensional Dirac Hamiltonian:
\begin{equation}\label{sdh}
\mathcal{H}_s= iv_F{\boldsymbol  \sigma} \cdot \textbf{k} 
\end{equation}
where  ${\boldsymbol  \sigma}=(\sigma_x,\sigma_y)$ are the Pauli matrices in the bases  $\left(\psi_\uparrow,\psi_\downarrow\right)$, and $\psi_\sigma$ is the electronic state with spin $\sigma$ localized on the surface of the TI.

The low-energy effective Hamiltonian describing a Josephson junction on the surface of the TI with supercurrent along $\hat{x}$ (such that the superconducting $\phi$ varies in   $\hat{x}$-direction) is given by\cite{PhysRevLett.107.097001}
\begin{equation} \label{sh}
H=\left(-iv_F{\boldsymbol{\nabla}}\cdot\boldsymbol{\sigma}-\mu\right)\eta_z+\Delta\left[\cos(\phi(x))\eta_x+\sin(\phi(x))\eta_y\right]\ ,
\end{equation} 
with a convenient choice of bases  $\left(\psi_\uparrow,\psi_\downarrow,\psi_\downarrow^\dagger,-\psi^\dagger_\uparrow\right)^T$. In this Hamiltonian, the Fermi velocity at chemical potential $\mu$ is denoted by $v_F$, and $\Delta$ is the superconducting gap.  The matrices $\boldsymbol{\sigma}$ act on physical spin space (which is locked with the momentum) whereas the $\eta_i$ act on the superconducting particle-hole space.

As the Hamiltonian, eqn.~\ref{sh}, is invariant under translation along $\hat{y}$, the momentum  $k_y$ in this direction is conserved. The lowest-energy Andreev states in the junction correspond to $k_y=0$ for which equation \ref{sh} reduces to:
\begin{equation} \label{Hamil}
\left[\left(-iv_F\partial_x\sigma_x-\mu\right)\eta_z+\Delta\left(\cos(\phi(x))\eta_x+\sin(\phi(x))\eta_y\right)\right]|v\rangle=E|v\rangle\ .
\end{equation}
As an operator, $\sigma_x$ commutes with the above Hamiltonian. Therefore, it is possible to divide the eigenstates into two separate sets with eigenvalue $\langle\sigma_x\rangle=1$, or $-1$:
\begin{equation}
\left[\left(\mp iv_F\partial_x-\mu\right)\eta_z+\Delta\left(\cos(\phi(x))\eta_x+\sin(\phi(x))\eta_y\right) \right]|v\rangle=E|v\rangle\ ,
\end{equation}
where the negative (positive) sign of the first term corresponds to $\langle\sigma_x\rangle = 1$ (or $-1$), respectively. 
Similarly, at finite, positive chemical potential ($\mu>0$), the right and left Fermi points correspond to states with opposite $\sigma_x$ eigenvalues -- $\langle\sigma_x \rangle > 0$ for the right, and  $\langle\sigma_x \rangle < 0$ for the left Fermi point, respectively.

The eigenvectors $|v\rangle$ have the form $|v\rangle=e^{i k_\pm x}|u\rangle$, where $k_\pm=\pm k_f= \pm
\frac{\mu}{v_F}$ are the Fermi momenta at the two Fermi points, and the eigenvectors $|u \rangle$ satisfy:
\begin{equation}\label{Hamils}
\left[\left(\mp iv_F\partial_x\right)\eta_z+\Delta\left(\cos\left(\phi(x)\right)\eta_x+\sin\left(\phi(x)\right)\eta_y\right)  \right]|u\rangle=E|u\rangle\ .
\end{equation}

A short Josephson junction is modeled by the following phase distribution:
\begin{equation}
\left\{\begin{array}{cll}  \phi(x)=& 0 \; &\mathrm{for} \  x<0,\ \mathrm{and} \\ \phi(x)=& \phi_0 \; &\mathrm{for} \  x\ge 0\ . \end{array}\right.
\end{equation}
Since we are interested in in-gap Andreev states, we define $E=\Delta \cos(\beta)$ with $0 \le \beta \le \pi$. The eigenvector $|u\rangle$ has the form: 
\begin{equation}
|u\rangle=e^{\frac{\kappa \Delta}{v_F}  x}\left(\begin{array}{ccc} a \\ b \end{array} \right)
\end{equation}
where the vectors $\left(a,b\right)^T$ satisfy the following equation:
\begin{equation}
\Delta\left(\begin{array}{ccc} \mp i\kappa-\cos(\beta) & e^{i\phi(x)} \\  e^{-i\phi(x)} & \pm i\kappa-\cos(\beta) \end{array}\right)\left(\begin{array}{ccc} a \\ b \end{array} \right)=0\ ,
\end{equation}
which leads to $\kappa=\sin(\beta)$ for $x<0$, and  $\kappa=-\sin(\beta)$ for $x>0$. Notice, since we have $0 \le \beta \le \pi$, it follows $\sin(\beta)\ge 0$. The wave functions are:
\begin{equation}
x<0:\ \left(\begin{array}{ccc} a \\ b \end{array} \right)=\frac{1}{\sqrt{2}} \left(\begin{array}{ccc} 1  \\ e^{\pm i\beta} \end{array} \right)\ ,\ \mathrm{and}
\end{equation}
\begin{equation}
x\ge 0:\ \left(\begin{array}{ccc} a \\ b \end{array} \right)=\frac{1}{\sqrt{2}} \left(\begin{array}{ccc} 1 \\ e^{i(\phi \mp \beta)} \end{array} \right)\ .
\end{equation}

From the continuity condition at $x=0$, we obtain $\beta=\frac{\phi}{2}$ for the right Fermi point, and $\beta=\pi-\frac{\phi}{2}$ for the left Fermi point such that $0 \leq \phi,\ \beta \leq \pi$. Close to right and left Fermi points, the energies of the ABS are found to be $E=\Delta \cos(\phi/2)$ and $E=\Delta\cos(\pi-\phi/2)=-\Delta\cos(\phi/2)$, respectively. The two modes correspond to eigenvalues of $\sigma_x$ with opposite sign, thus, there is no mixing between them. Importantly, their wave functions stay localized on the surface of the TI because bulk ABS have larger energy. In what follows, we show that the energy of Andreev states in the bulk of a TIJJ is always larger than the energy of low-energy Andreev states on the surface, unless the chemical potential corresponds to the Fermi wave vectors $\textbf{k}_f$ where the effective mass  $|m(\textbf{k}_f)|$ vanishes.

The low-energy Hamiltonian in the bulk of a topological insulator is the massive Dirac Hamiltonian, eqn.~\ref{bh}. Adding terms for superconducting pairing and the finite chemical potential, the Hamiltonian reads as
\begin{equation}
H_{b}=\left[v_F\left( {\boldsymbol{\nabla}}\cdot\boldsymbol{\sigma} \right) \tau_x+m \left(|\textbf{ k}| \right) \tau_z - \mu \right] \eta_z + \Delta\left[\left(\cos\left(\phi(x)\right)\eta_x+\sin\left(\phi(x)\right)\eta_y\right)  \right]\ .
\end{equation}
Here, we would like to point out that the minimal model for TI bulk states has two orbitals, and another set of Pauli matrices, $\tau_i$, for orbital space had to be introduced in the Hamiltonian. Therefore, the algebraic structure of the wave function is different for bulk and surface states. 

Similar to surface states, low-energy bulk ABS have momenta parallel to the $\hat{x}$-direction (i.e. $k_y=k_z=0$), and the effective Hamiltonian is
\begin{equation}
H_{b}=\left[ v_F k_x \sigma_x  \tau_x+m \left(|\textbf{k}| \right) \tau_z - \mu \right] \eta_z +\Delta\left[\left(\cos\left(\phi(x)\right)\eta_x+\sin\left(\phi(x)\right)\eta_y\right)  \right]\ .
\end{equation}
Again, the operator $\sigma_x$ commutes with the effective Hamiltonian thus we can divide the eigenstates into two separate sets with eigenvalue $\langle\sigma_x\rangle=1$, or $-1$:
\begin{equation}\label{Hamilb}
H_{b}=\left[\pm v_F k_x \tau_x+m \left(|\textbf{ k}|\right) \tau_z - \mu \right] \eta_z + \Delta\left[\left(\cos\left(\phi(x)\right)\eta_x+\sin\left(\phi(x)\right)\eta_y\right)  \right]\ .
\end{equation}

In the following, we only present the solution for $\langle\sigma_x\rangle=1$, which corresponds to the $+$ sign for the first term of $H_{b}$. The case of $\langle\sigma_x\rangle=-1$ can be treated identically. When the chemical potential is in the conduction band, spectrum and orbital wave function are solutions of the kinetic Hamiltonian, $\mathcal{H}=v_F \tau_x k_x + \tau_z \ m(\textbf{k})$.  Its eigenvalues and eigenfunctions are given by:
\begin{equation}\label{wf}
\begin{split}
\mathcal{E}(k_x)&=\sqrt{v_F^2 k_x^2+ m \left(|k_x|\right) ^2}\ , \\
|\Phi_\tau(k_x)\rangle &=\frac{1}{\mathcal{N}} \left(m(|k_x|),\sqrt{m(|k_x|)^2+v_F^2 k_x^2}-v_F k_x\right)^T\ ,
\end{split}
\end{equation}
with the normalization factor $\mathcal{N}^2=2m(|k_x|)^2+2v_F^2 k_x^2-2v_F k_x\sqrt{m(|k_x|)^2+v_F^2 k_x^2}$. The low-energy ABS close to the Fermi points, $k^{\pm}_f=\pm \frac{\sqrt{\mu^2-m \left(|k_f|\right) ^2}}{v_F}$, are derived by setting $k_x=k^{\pm}_f\mp i\partial_x$ in equation \ref{Hamilb} and by projecting onto the corresponding orbital wave functions $|\Phi_\tau(k^\pm_f)\rangle$. We obtain the equations
\begin{equation}\label{fp}
\left[\left(\mp iv_F\partial_x\right)\eta_z+\Delta\left(\cos(\phi(x))\eta_x+\sin(\phi(x))\eta_y\right) \right]|w\rangle=E|w\rangle
\end{equation}
The Hamiltonian in eqn.~\ref{fp} is formally identical to the previously discussed Hamiltonian, eqn.~\ref{Hamils}. For the energies at the right and left Fermi points, we have $E=\Delta \cos(\phi/2)$ and $E=\Delta\cos(\pi-\phi/2)=-\Delta\cos(\phi/2)$. 
Unlike before, the wave functions at the two Fermi points correspond to the same eigenvector of $\sigma_x$, and the orbital parts of the wave functions, eqn.~\ref{wf}, are -- in general -- not orthogonal. Since the phase $\phi$ varies along the $\hat{x}$-direction, states at the two Fermi points are scattered into each other and repel.

The energy shift due to scattering is calculated using second-order perturbation theory. It is proportional to the orbital overlap of the wave functions at the two Fermi points, i.e. $\langle \Phi_\tau(k^-_f)|\Phi_\tau(k^+_f)\rangle =  \frac{| m \left(|k_f|\right)|}{\sqrt{m \left(|k_f|\right)^2+v_F^2 k_f^2}}$, and takes the form
\begin{equation}
E=\pm\left( \Delta |\cos(\phi/2)|+\alpha \frac{m \left(|k_f|\right)^2}{m \left(|k_f|\right)^2+v_F^2 k_f^2}\right)
\end{equation}
where $\alpha>0$. Assuming the magnitude of the superconducting gap is equal on the surface and in the bulk, it follows that the energies of bulk Andreev states are larger than those of surface Andreev states unless $m\left(\mathbf{k}_f\right)=0$.

\ 

\textbf{Numerical simulation}

In order to further support our model, we study the structure of low-energy Andreev states in TIJJs numerically. A simple discrete model for TIs includes 4 orbitals on a cubic lattice with orbital-dependent nearest-neighbor hopping\cite{PhysRevB.81.045120, PhysRevLett.107.097001}. We implement superconductivity at the mean-field level, i.e. we double the number of orbitals at each lattice site to represent superconducting particle- and hole-states, and we add a coupling term between the two sectors. Allowing the phase of this coupling to vary along the $\hat{x}$-direction, we can model a TIJJ. 

As we set the momentum along the $\hat{y}$-direction (i.e. ``parallel'' to the junction) equal to zero, it is sufficient to discretize the Hamiltonian on a square, real-space lattice with dimensions $L_x \times L_z$ for which the two site-labels~$(x,z)$ are chosen along the directions of the superconducting phase variation, $\hat{x}$, and the chemical potential shift, $\hat{z}$. Band-bending on the surface of the TI is modeled by choosing the chemical potential for sites with index $0<z \le L_z/3$ in the conduction band whereas it is fixed at zero for sites with $z>L_z/3$. We then solve the discrete Hamiltonian exactly and plot the wave function amplitudes associated with the lowest-energy ABS against the chemical potential in the surface layer, see Fig.~3(f) in the main article. We observe that, as the chemical potential enters the conduction band, most of the weight of the ABS wave functions remains close to the boundary of the model (i.e. at $L_z = 0$). When we further increase the chemical potential, the wave function spreads gradually across the surface region. As the critical chemical potential ($\mu_C$) is reached, and the effective mass vanishes ($m\left(\textbf{k}_f\right)=0$), the wave function is fully delocalized and extends throughout the region (i.e. sites for which $0<z<L_z/3$) where band-bending occurs in our model. Past the critical value $\mu_C$, the ABS are strongly localized at $z=L_z/3$, which is the interface between the -- now --  topologically-trivial surface and the band-inverted bulk.

In conclusion, we showed that in a doped topological insulator, where superconductivity is induced in the bulk (in addition to the surface states), the energies of low-energy bulk Andreev-bound states are related to the magnitude of the effective mass of the bulk band and have a minimum when the effective mass vanishes. At this point, the energies of boundary and bulk ABS are equal, which couples them and annihilates the boundary modes. Most prominently, this mechanism destroys surface Majorana modes when the bulk substrate undergoes a topological transition which makes it relevant in a wider context. Experimental signatures of the spatial displacement of boundary ABS due to a topological transition in the bulk band are covered in the main text.   

\end{addendum}

\end{document}